\documentclass[11pt,a4paper]{article}
\usepackage{epsfig,cite}
\newcommand{\gstar}{\gamma^\ast }
\newcommand{\comment}[1]{}
\newcommand{\aver}[1]{\langle#1\rangle}

\newcommand{\degr}{^\circ}
\newcommand{\lsim}{~\raisebox{-0.5mm}{$\stackrel{<}{\scriptstyle{\sim}}$}~}
\newcommand{\gsim}{~\raisebox{-0.5mm}{$\stackrel{>}{\scriptstyle{\sim}}$}~}
\newcommand{\xpom}{x_{I\!\!P}}
\newcommand{\pom}{I\!\!P}
\newcommand{\avg}[1]{\langle{#1}\rangle}

\begin{document}
\bibliographystyle{unsrt}

\begin{titlepage}

\begin{flushleft}

{\tt DESY 98-044\hfill ISSN 0418-9833} \\
\end{flushleft}
\vspace*{2.5cm}
\begin{center}
\begin{LARGE}

{\bf Multiplicity Structure of the Hadronic Final State in
Diffractive Deep-Inelastic Scattering at HERA}

\vspace*{2cm}

H1 Collaboration \\

\vspace*{2.5cm}
\end{LARGE}

{\bf Abstract}

\begin{quotation} 
\noindent The multiplicity structure of the hadronic system $X$
produced in deep-inelastic processes at HERA of the type $ep
\rightarrow eXY$, where $Y$ is a hadronic system with mass 
$M_Y< 1.6$~GeV and where the squared momentum transfer at the $pY$
vertex, $t$, is limited to $|t|<1 {\rm\ GeV}^2$, is studied
as a
function of the invariant mass $M_X$ of the system $X$.  
%
Results are presented on multiplicity
distributions and multiplicity moments, rapidity spectra
and forward-backward correlations in the centre-of-mass system of $X$.
The data are compared to results in $e^+e^-$ annihilation,
fixed-target lepton-nucleon collisions, hadro-produced diffractive
final states and to non-diffractive  hadron-hadron collisions.
The comparison suggests  a production mechanism of
virtual photon dissociation which involves a mixture of
partonic states and  a significant gluon content.
The data are well described by a model, based on a QCD-Regge analysis
of the diffractive structure function, which assumes a
large hard gluonic component of the colourless exchange at low $Q^2$.
A model with soft colour interactions is also successful.


\end{quotation}
\vspace*{1cm}
\end{center}
\vspace*{3cm}
\begin{center}
\em Submitted to Eur.~Phys.~J.
\end{center}

\cleardoublepage
\end{titlepage}

\noindent C.~Adloff$^{34}$,                
 M.~Anderson$^{22}$,              
 V.~Andreev$^{25}$,               
 B.~Andrieu$^{28}$,               
 V.~Arkadov$^{35}$,               
 I.~Ayyaz$^{29}$,                 
 A.~Babaev$^{24}$,                
 J.~B\"ahr$^{35}$,                
 J.~B\'an$^{17}$,                 
 P.~Baranov$^{25}$
% ...

\vspace*{0.6cm}
\par\noindent
% ...
$ ^{17}$ Institute of Experimental Physics, Slovak Academy of
          Sciences, Ko\v{s}ice, Slovak Republic$^{f,j}$ \\
% ...
$ ^{22}$ Department of Physics and Astronomy, 
          University of Manchester, Manchester, UK$^b$ \\
% ...
$ ^{24}$ Institute for Theoretical and Experimental Physics,
          Moscow, Russia \\
 $ ^{25}$ Lebedev Physical Institute, Moscow, Russia$^{f,k}$ \\
% ...
$ ^{28}$ LPNHE, Ecole Polytechnique, IN2P3-CNRS, Palaiseau, France \\
 $ ^{29}$ LPNHE, Universit\'{e}s Paris VI and VII, IN2P3-CNRS,
          Paris, France \\
% ...
$ ^{34}$ Fachbereich Physik, Bergische Universit\"at Gesamthochschule
          Wuppertal, Wuppertal, Germany$^a$ \\
 $ ^{35}$ DESY, Institut f\"ur Hochenergiephysik, Zeuthen, Germany$^a$ \\
% ...
$ ^b$ Supported by the UK Particle Physics and Astronomy Research
       Council, and formerly by the UK Science and Engineering Research
       Council \\
% ...
$ ^f$ Supported by the Deutsche Forschungsgemeinschaft \\
% ...
$ ^j$ Supported by VEGA SR grant no. 2/1325/96 \\
 $ ^k$ Supported by Russian Foundation for Basic Researches 
       grant no. 96-02-00019 \\

\newpage


\section{Introduction}

The observation of ``Large Rapidity Gap'' (LRG) events
\cite{bib:zeus_diff,bib:h1_diff}~ in deep-inelastic $ep$
scattering (DIS) at HERA, which are mainly attributed to
diffractive photon dissociation\cite{bib:h1:diff1,bib:zeus:diff1},  
has  led to renewed interest in diffractive phenomena~\cite{dd:reviews} 
and how they can be
understood within quantum chromodynamics (QCD).
This  paper continues a series of final-state studies by
H1~\cite{h1:flow96,bib:h1:mxpaper,bib:h1:thrust,bib:h1:finalstates}
and ZEUS~\cite{zeusddpaper1,zeusddpaper2,bib:zeus_eventshape} 
and presents the first results on  the charged particle multiplicity
structure in  LRG events.

Following conventions  adopted in earlier H1
analyses~\cite{bib:h1:mxpaper,bib:h1:f23d},
the  large rapidity gap events studied here  are
experimentally defined in
terms of the generic process $ep \rightarrow eXY$, where the
hadronic systems $X$ and $Y$ are separated by the largest rapidity
gap in the event; $Y$ is the system closest to the
outgoing proton beam \cite{bib:h1:mxpaper,bib:h1:f23d,bib:h1:thrust}.
Events
with no activity in a large pseudo-rapidity\footnote{H1 uses a
laboratory coordinate system with the $z$-axis aligned with the proton
beam direction.  Pseudorapidity is defined as
$\eta=-\ln(\tan\frac{\theta}{2})$, where $\theta$ is the polar
angle with respect to the proton direction.} 
domain adjacent to
the outgoing proton beam are selected.  In these events the
proton either stays intact ($Y =$~proton), or is excited to a
low-mass system. 
The system $Y$  has longitudinal momentum close to that of 
the proton beam and small transverse momentum.
%
%
In such events, the system  $X$, measured in the central
part of the detector, can be viewed as mainly resulting from the
dissociation of a  photon with a virtuality $Q^2$.
The requirement of a large rapidity gap  implies that the invariant
masses $M_X$ and $M_Y$ of the systems $X$ and $Y$ are small compared to
$W$, the centre-of-mass energy of the $\gstar p$ system. 
In the following, the terms ``LRG events'' and ``diffractive events'' will
be used synonymously, although, in practice,  some contribution from
non-diffractive and  double-diffractive processes  is to be
expected, particularly at large $M_X$.

In diffractive DIS it is useful to view the interaction in
a Lorentz frame where the target proton is at rest.
For small values of Bjorken-$x$, $x_{Bj}$,  the  virtual photon
fluctuates far upstream of the proton target into a
quark-antiquark pair which can subsequently evolve into a more complex
partonic system ($q\bar q,q\bar q g,\ldots$) before the actual interaction
occurs~\cite{bib:ioffe,bjorken:kogut,nik:zak:91}.
The  virtual photon can thus be
described  as a superposition of Fock-states with different partonic
content~\cite{abramowicz}.

Diffraction in general~\cite{good:walker}, and dissociation of the photon
in
particular, arises from the fact
that the strength of absorption of the various Fock-states
depends on the internal degrees of freedom and quark-gluon  composition
of the
dissociating object~\cite{miett:pumplin,blattel,bertsch}.
Specifically,  whereas the total $\gstar p$ cross
section measures the average absorption strength,
the magnitude  of the diffractive
cross section, on the other hand, is related to  its
fluctuations~\cite{miett:pumplin,blattel}.
The diffractively produced hadronic final states are therefore expected
to carry
 information, not only  on the parton composition   of the virtual 
photon wavefunction components and their
interaction with the target, but also on their respective contributions
to the diffractive cross section.




Virtual photon dissociation can also be approached from a $t$-channel
perspective.  In Regge phenomenology, the interaction takes place
via a factorisable exchange of the pomeron ($\pom$)  and of
reggeons related to mesons.  It has been suggested to endow the
pomeron (and reggeons) with a partonic sub-structure and to use
the concept of parton distributions in the pomeron and in
sub-leading reggeons to model diffractive deep-inelastic
scattering \cite{pom}. In the proton's infinite-momentum frame,
the pomeron (or reggeon) has a fraction $\xpom = (q \cdot (P-P')) /
(q \cdot P)$ of the proton's four-momentum $P$ ($q$ and $P'$ being the
four-momentum of the exchanged photon and the system $Y$,
respectively), while the fractional momentum of the exchange
carried by the struck parton is $\beta = x_{Bj} / \xpom$.  
 This approach, adopted in \cite{bib:h1:f23d} assuming
the DGLAP evolution equations~\cite{dglap}, indicates, as was already
conjectured in~\cite{low:nussinov}, that the pomeron must have a
large hard gluon content at low $Q^2$  but that a sub-leading meson
exchange is also needed at larger values of $\xpom$ or $M_X$, if the
basic hypothesis of factorisation of each component is to be
maintained. 

The parton structure of the colourless exchange, as deduced from an
analysis of the total diffractive DIS cross
section, can be tested in studies of
diffractive final
states~\cite{bib:h1:thrust,bib:zeus_eventshape,bib:h1:finalstates,bib:h1_diffjets}. 
These   confirm the need for a pomeron dominated by
hard gluons at a starting scale $Q_0^2\sim3$ GeV${}^2$.

This paper complements previous work and presents  an analysis of
charged particle multiplicity distributions, multiplicity 
moments, charged particle density in rapidity space
and forward-backward multiplicity
correlations measured in the centre-of-mass (CMS) of the system
$X$. 
The emphasis is placed on a
comparison with data from   $e^+e^-$ annihilation,
fixed-target DIS, hadron diffraction and  soft non-diffractive
hadron-hadron collisions.
Monte Carlo models 
which represent various theoretical views on diffraction 
in DIS~\cite{lepto64,bib:rapgap} are also confronted with the data.


\section{Experimental procedure}

\subsection{The experiment}

The experiment was carried out with the H1 detector
\cite{bib:h1_detector} at the HERA storage ring at DESY. The data
were collected during the 1994 running period when 27.5 GeV
positrons collided with 820 GeV protons, at a centre-of-mass
energy of 300 GeV. The following briefly describes the detector
components most relevant to this analysis. 

The energy of the scattered positron is measured with a backward
electromagnetic lead-scintillator calorimeter (BEMC), extending
over the polar angular range $151\degr<\theta<177\degr$ with full
azimuthal coverage. The BEMC electromagnetic energy resolution is
$\sigma_E/E \approx 0.10/\sqrt{E [{\rm GeV}]}$ $ \oplus~0.42/E
[{\rm GeV}] \oplus 0.03$, while the BEMC energy scale for
positrons is known to an accuracy of 1\% \cite{bib:h1_bemc}. A
backward proportional chamber (BPC), situated immediately in front
of the BEMC and with an angular acceptance of $155.5\degr < \theta
< 174.5\degr$, serves to measure the impact point of the scattered
positron and to confirm that the particle entering the BEMC is
charged.  Using information from the BPC, the BEMC and the
reconstructed event vertex, the polar angle of the scattered
positron can be determined to better than 1 mrad.  A scintillator
hodoscope behind the BEMC is used to reject beam-induced
background based on a time-of-flight measurement.

The hadronic final state is measured by tracking detectors
surrounded by calorimeters.  The Central Tracker consists of inner
and outer cylindrical jet chambers, $z$-drift chambers and
proportional chambers.  The jet chambers, mounted concentrically
around the beam line, inside a homogeneous magnetic field of 1.15~Tesla,
provide both particle charge and momentum measurement from
track curvature and cover an acceptance region defined by the
angular interval $15\degr < \theta < 165\degr$ and transverse
momentum $p_T > 0.1$ GeV. Up to 56 space points can be measured
for tracks with sufficiently large $p_T$.  The resolutions
achieved are $\sigma_{p_T}/p_T \approx 0.009 \cdot p_T {\rm\
[GeV]} \oplus 0.015$ and $\sigma_{\theta} = 20$ mrad
\cite{bib:h1_pt,bib:h1_detector} with a track finding efficiency
above 95\% for tracks well contained in both jet chambers.  In addition,
forward going particles can be detected
by the Forward Tracker in the polar angular range $8\degr < \theta
< 20\degr$.  The liquid argon calorimeter (LAr)  extends over the
polar angular range $4\degr < \theta < 154\degr$ with full
azimuthal coverage.  The LAr hadronic energy resolution is
$\sigma_E/E \approx 0.50/\sqrt{E [{\rm GeV}]} \oplus 0.02$ as
determined in test beams \cite{bib:h1_calo}.  A study of the
transverse momentum balance between the hadronic final state and
the scattered positron has shown that the absolute hadronic energy
scale is known to an accuracy of 4\%.  Backward going hadrons can
be detected by the BEMC.  The hadronic energy scale of the BEMC is
known to a precision of 20\% \cite{bib:h1_bemc}.

Forward energy at small angles is observed in several detectors
near the outgoing proton beam direction.  Particles reach these
detectors both directly from the interaction point and as a result
of secondary scattering with the beam pipe  and other adjacent
passive material.  The detectors are thus sensitive to
energetic particles produced in directions that are  beyond their
geometrical acceptances.  The effective ranges of sensitivity to energy
flow are 
$3.5 \lsim \eta \lsim 5.5$ for
the copper/silicon sandwich (PLUG) calorimeter, $4.5 \lsim \eta
\lsim 6.5$ for the Forward Muon Spectrometer and $6.0 \lsim \eta
\lsim 7.5$ for the Proton Remnant Tagger, which consists of
scintillation counters and is located 24 m from the interaction
point~\cite{bib:h1:diff1}.  These detectors overlap considerably in their
rapidity coverage, thereby allowing for intercalibration of their
efficiencies. 

\subsection{Monte Carlo simulation of LRG events}

Monte Carlo generated LRG events obtained with the RAPGAP~2.02
generator~\cite{bib:rapgap} are used to correct the observed
distributions for detector acceptance, inefficiencies and smearing
effects. This model correctly accounts for many final-state
features of LRG
events~\cite{bib:h1:f23d,bib:zeus_eventshape,bib:h1:thrust,bib:h1:finalstates},
but was not tuned to the multiplicity data presented in this
paper.  All generated events go through a full simulation of the
H1 apparatus and are passed through the same analysis chain as the
real data. The detector simulation is based on the GEANT
program~\cite{bib:geant}.  The RAPGAP predictions shown in 
subsequent figures were derived from a model version which
incorporates the recent results of the QCD-Regge analysis of the
diffractive structure function~\cite{bib:h1:f23d}. 
%
Results are also shown from the DIS Monte Carlo event generator
LEPTO~6.5~\cite{bib:lepto65} and from the JETSET
parton-shower model~\cite{bib:jetset} for  simulation of $e^+e^-$
annihilation hadronic final states.  Important technical aspects of these
models
are given in recent H1 
publications~\cite{bib:h1:f23d,bib:h1:thrust,bib:h1:finalstates} 
and are not repeated here. 

The RAPGAP generator  models LRG events as
deep-inelastic scattering of a virtual photon off a pomeron or
reggeon coupled to the initial-state proton.  
The pomeron and
reggeon  are given a partonic content.  The meson
structure function is taken to be that of the pion
\cite{pionpdf:owens}.

Different partonic sub-processes are implemented
using Born-term and first-order perturbative QCD matrix elements: $e q
\rightarrow
e q$ scattering, QCD-Compton scattering ($e q \rightarrow e q g$) 
and boson-gluon fusion (BGF) off a gluon in the colourless
exchange ($e g \rightarrow e q \overline{q}$). Their relative
contributions are controlled by quark and gluon densities of the
exchange  as determined in the H1 QCD-Regge analysis
\cite{bib:h1:f23d} of the LRG event  cross section using the DGLAP
evolution equations. In this model a ``pomeron remnant'' heads in
the direction opposite to the virtual photon, consisting of a
quark or a gluon for $eq$ or $eg$ scattering, respectively.
The fragmentation of the partonic systems created in the
sub-processes  $e q\rightarrow e q$, $e q \rightarrow e q g$, is
thus expected
to be analogous to that in $e^+e^-$ annihilation with
centre-of-mass energy $\sqrt{s}=M_X$. However, in BGF
the  initial partonic system consists of
a pomeron remnant (gluon)  and a $q\overline{q}$ pair in a colour-octet
state. This process has no equivalent in $e^+e^-$ annihilation.

To assess the sensitivity to the quark-gluon
content of the pomeron, results are presented for two sets of
parton distributions (labelled hereafter ``RG~$F_2^D$~(fit~3)'' and
``RG~$F_2^D$~(fit~1)''):  i) a ``hard gluon''
distribution (``fit~3'' in~\cite{bib:h1:f23d}) whereby gluons
carry $\geq80\%$ of the momentum at the starting  scale
$Q_0^2=3$~GeV${}^2$; ii) a ``$q\overline{q}$-only'' distribution
whereby only quarks are present at $Q_0^2$ (``fit~1''
in~\cite{bib:h1:f23d}). The latter model version is disfavoured
in fits to  $F_2^{{D}(3)}$~\cite{bib:h1:f23d} and  by
diffractive final-state studies reported
in~\cite{bib:zeus_eventshape,bib:h1:thrust,bib:h1:finalstates}\footnote{The
predictions from fit~2 in~\cite{bib:h1:f23d} are very  similar to those of
fit 3 for the observables studied here and not shown.}. 

Higher-order effects in the QCD cascade are treated with the
parton shower model, as implemented in LEPTO \cite{bib:lepto}.
Hadronisation is carried out with the Lund string fragmentation
scheme, as in JETSET~7.4~\cite{bib:jetset}. QED radiative
processes are included via an interface to the program HERACLES
\cite{bib:heracles}.

Soft colour interactions form  the basis of an alternative model,
which is implemented in the LEPTO~6.5~\cite{lepto64} generator. In
this scheme~\cite{sci:ingelman}, LRG events are the
result of a normal deep-inelastic  scattering on the proton followed by a
long-time-scale random colour rearrangement in the soft
field of the target. This mechanism leads to a fraction of events
with a large rapidity gap of the type studied in this paper.

Studies of  diffractively produced meson systems  
in meson-hadron interactions
with masses $M_X$ below $10$~GeV 
 have revealed striking
similarities with $e^+e^-$
final states at $\sqrt{s}=M_X$~(see
e.g.~\cite{bib:na22_diffmult}).  To analyse virtual photon
dissociation along similar lines, the JETSET parton shower model
for this process is used (labelled hereafter ``JETSET-$e^+e^-$''). 
The predictions are calculated for a standard
mixture of primary ($u,d,s$) quark-antiquark pairs only. 
A primary  $c\overline{c}$ component ($e^+e^-\rightarrow c\bar c$)  has
been neglected,
motivated by the prediction that, for a  quark-dominated pomeron, 
heavy-quark production is 
suppressed~\cite{brodsky:rapgap}.
It was verified, however,
that  inclusion of  a contribution from primary  $c\bar c$ pairs
has little effect on the multiplicity and would not alter
the  conclusions of this analysis. The generated $e^+e^-$ events are
rotated
such that the $z$-axis coincides with the initial $q\overline{q}$
axis of the event in the CMS. Rapidity\footnote{
Rapidity of a particle is defined as $y = 0.5
\ln\left[(E+p_z)/(E-p_z)\right]$, where $E$ is the energy and
$p_z$ the momentum component along the direction of a predefined  axis;
the pion mass is assigned to each particle.} is
calculated relative to
that direction. The use of JETSET-$e^+e^-$ permits a
consistent comparison with the H1 data and avoids difficulties
due to the differing experimental treatment of e.g.~strange particle production
in different $e^+e^-$ experiments.

To study the systematic uncertainties arising from background
processes, several other Monte Carlo generators are used. 

The DIFFVM generator~\cite{bib:diffvm} models the low mass region
of diffraction ($M_X < 1.1$ GeV) via production of the 
vector mesons $\rho(770)$,
$\omega(782)$ and $\phi(1020)$.  This model further includes a
simulation of proton diffractive dissociation. 

The PHOJET generator~\cite{bib:phojet} is used to estimate
background from photoproduction processes. The model simulates
non-diffractive reactions, elastic vector meson production, vector
meson production with proton dissociation, single-photon
diffractive dissociation and double diffraction and is in broad agreement
with experimental results at
HERA~\cite{bib:h1_photoproduction}.

\subsection{Event and track selection}

A neutral current DIS event selection is made~\cite{bib:h1_multiplicity} 
by demanding a
well-reconstructed scattered positron detected in the BEMC with an
energy, $E_e'$, larger than 10 GeV.  A subsample of dominantly
diffractive events is then selected by requiring no activity above
noise levels in any of the forward detectors or in the most
forward part ($\eta > 3.2$) of the LAr Calorimeter.

Further cuts are applied to ensure that a positron is detected and
reconstructed with high quality.  An event vertex, reconstructed
from tracks in the central trackers, within $z=\pm 30$ cm of the
mean vertex position, is required to reject beam-induced
background.  Events with a time-of-flight veto from the
scintillator hodoscope are rejected. 

The standard deep-inelastic kinematic variables ($y_{Bj}$, $Q^2$) 
are reconstructed with the methods described
in~\cite{bib:h1:f23d,bib:h1:mxpaper}
and which  use  both 
the scattered positron and the hadronic final state. The kinematic
variables ($\xpom$, $M_X$) characterising the final state in LRG
events, are obtained from a combination of tracker and calorimeter
information with an algorithm for track-cluster association which
avoids double counting. In~\cite{bib:h1:f23d,bib:h1:mxpaper} 
it is demonstrated that $M_X$ is adequately
reconstructed across the kinematic range of the measurement with a
resolution of about $25\%$.

Corrected data are restricted to a kinematic region where the
acceptance of the H1 detector is high and the contribution from
non-DIS background  is low.  Together with the
requirement $E_e' > 10$ GeV, an upper limit on $y_{Bj}$ at 0.6
ensures that the photoproduction background is less than 0.3\% in
the selected event sample. The lower limit $y_{Bj} > 0.05$ ensures
substantial hadronic energy flow in the detector and adequate resolution
in  $y_{Bj}$.
The
$y_{Bj}$ cuts correspond roughly to a range $70 < W < 230$
GeV.  Non-diffractive
contributions are suppressed by requiring $\xpom < 0.05$. 
A lower cut on $M_X$ of 3~GeV
excludes  light vector meson contributions. In addition,
the requirement of an absence of activity in the forward detectors
imposes the approximate restrictions $M_Y \lsim 1.6\ {\rm GeV}$
and $|t| \lsim 1\ {\rm GeV}^2$, although these variables were not
measured directly. The kinematic regions to which the data are
corrected are summarised in Table~\ref{tab:kinematics}; there  $|t_{min}|$ 
is the minimal kinematically allowed value of  $|t|$. 
The event sample consists of 4738 events, corresponding to
an integrated luminosity of $1.3\ {\rm pb}^{-1}$. 

\begin{table}[ht]
\begin{center}
\begin{tabular}{c|c|c}
Quantity    & Lower limit        & Upper limit        \\ \hline
$Q^2$       & $7.5\ {\rm GeV^2}$ & $100\ {\rm GeV^2}$ \\
$y_{Bj}$    & 0.05               & 0.6                \\
$\xpom$     & 0.0003             & 0.05               \\
$M_X$       & 3 GeV              & 36 GeV             \\
$|t|$       & $|t_{min}|$        & $1\ {\rm GeV^2}$   \\
$M_Y$       & proton mass        & 1.6 GeV            \\
\end{tabular}
\caption{Limits of the kinematic regions considered for
corrected data.}
\label{tab:kinematics}
\end{center}
\end{table}

The multiplicity analysis is based on charged particles.  Only
tracks observed within the acceptance of the central tracking
detector, and which are successfully fitted to the primary event
vertex, contribute to the uncorrected multiplicity. Further
details on the track selection criteria and efficiencies can be
found in~\cite{bib:h1_multiplicity}.

The multiplicity distribution is corrected with an iterative
matrix migration method based on full Monte Carlo simulation of
the detector response. The method is described in detail
in~\cite{bib:h1_unfolding,bib:h1_multiplicity}. The results are
cross-checked with a fit of a Negative Binomial distribution,
smeared for detector acceptance and inefficiencies, to the
observed multiplicity distribution.  This parametric technique is
known to be less sensitive to the generator input, but yields
only the lowest moments of the multiplicity distribution. For the
measurement of rapidity spectra, a standard bin-by-bin correction
procedure is used, with cross-checks provided by the matrix-unfolding
methods.

Charged decay products of $K^0_S$, $\Lambda$ and
$\overline{\Lambda}$ and from weakly decaying particles with
lifetimes larger than $8 \cdot 10^{-9}\ {\rm s}$ are subtracted
from the multiplicity distribution through the unfolding
method~\cite{bib:h1_multiplicity}.
Hadrons associated with the target remnant system $Y$ are
excluded
from the multiplicity measurement. 

\subsection{Systematic errors}
Several sources of possible systematic errors are investigated. 
The analysis is repeated for each source and the changes to the
results are added in quadrature.  For illustration, the typical 
systematic error on the mean total charged multiplicity and on the
central rapidity
($-0.5<y<0.5$) particle density are given in square
brackets for each source separately.

\begin{itemize}

\item The error due to the uncertainty of the energy
scale of the hadronic final state is
estimated by scaling the LAr, BEMC and Central Tracker
energies by $\pm 4\%$, $\pm 20\%$ and $\pm 3\%$ respectively
[0.4\%, 0.4\%].

The systematic uncertainty in the reconstruction of the
scattered positron is studied by varying the energy $E_e'$ and
polar angle $\theta'$ by $\pm 1\%$ and $\pm 1\ {\rm mrad}$,
respectively
[0.3\%, 0.9\%].

\item The influence of the pomeron (reggeon) flux and pomeron
(reggeon)  structure function used in the Monte Carlo generator
for correction is investigated by reweighing the $\beta$, $\xpom$
and $t$ distributions for Monte Carlo events as in~\cite{bib:h1:f23d} 
[$\beta$: 1.3\%, 2.2\%; $\xpom$: 0.3\%, 1.6\%; $t$: 1.1\%, 3.5\%].

\item Both the colour dipole model, as in ARIADNE \cite{bib:ariadne},
and the parton shower model are used to evaluate the influence of
these event generation schemes on the corrections.  The full difference
 is taken as the systematic error [0.6\%, 0.6\%]. 

\item The strangeness-suppression parameter ({\tt PARJ(2)} in
JETSET~\cite{bib:jetset}), affecting the rate of strange particle
production in the simulation of the fragmentation process, has been
varied in the range 0.2--0.3 according to recent results on
strange particle production \cite{delphi_strange,e665_strange}
[0.1\%, 0.1\%].

\item Track-quality criteria (such as the track length
and the number of hits) are varied to estimate systematic errors
related to an imperfect description of the acceptance and
efficiency of the Central Tracker in the Monte Carlo simulation
[1.3\%, 3.1\%]. 

An uncertainty of 30\% is assumed on the Monte Carlo correction
outside the tracker acceptance ($\theta_{LAB}$ outside the range
8$\degr$--165$\degr$ or $p^{lab}_T < 0.1$ GeV; the range between
8$\degr$--15$\degr$ has been cross-checked with data from the
Forward Tracker) [3.6\%, 1.5\%].


\item Background events are suppressed by the event selection
criteria.  Remaining background contamination is estimated by
including events simulated with the PHOJET and DIFFVM generators
in the Monte Carlo event sample [PHOJET: $< 0.1\%$, $< 0.1\%$; 
DIFFVM:  0.7\%, 1.0\%]. 

The number of events with initial and final-state QED radiation
is changed by $\pm 50\%$~[0.4\%, 0.1\%].

\item A fit to a smeared Negative Binomial distribution is used as
a cross-check on the matrix-unfolding results 
[error on central rapidity particle density 2.8\%].

\end{itemize}


\section{Results}

All data presented below\footnote{All data are available in
numerical form on request and can be retrieved from the Durham
HEPDATA database.} are corrected for the effects of acceptance and
resolution of the H1 detector in the kinematic ranges specified
in~Table~\ref{tab:kinematics}. 
To optimize the statistical precision of the different measurements
presented here,  a fine- and coarse-grained binning in $M_X$ is used.
The requirement of a forward
rapidity gap ensures that the hadronic final states of the system
$X$ are well contained in the central detectors. The data span the
$M_X$ range from 3 to 36 GeV, distributed over the intervals as
listed in Table~\ref{tab:intervals}. Statistical and systematic
errors on the data points shown in the figures are combined in
quadrature. Unless otherwise stated,  where two error bars are displayed
the inner one is
the statistical  and the  outer shows the total error.
In comparing H1 LRG data at a given $M_X$ to data from other
processes, the corresponding centre-of-mass energy scale is chosen to be
$W$ for fixed-target DIS data, $\sqrt{s}$ for $e^+e^-$ and non-diffractive
hadron collisions, $M_X$ for hadro-produced diffractive states.

\begin{table}[ht]
\begin{center}
\begin{tabular}{r@{--}l|r@{.}l|r@{.}l|c|c}
\multicolumn{2}{c|}{$M_X$ range} & 
\multicolumn{2}{c|}{$\avg{M_X}$} & 
\multicolumn{2}{c|}{$\avg{\beta}$} & 
\multicolumn{1}{c|}{$\avg{Q^2}$} &
no. of events \\  
\multicolumn{2}{c|}{(GeV)} &
\multicolumn{2}{c|}{(GeV)} &
\multicolumn{2}{c|}{ } &
\multicolumn{1}{c|}{${\rm GeV}^2$} & 
 \\ \hline
 3& 8 &  5&4 & 0&41 & 21 & 1492 \\
 8&15 & 11&4 & 0&17 & 26 & 1515 \\
15&30 & 21&1 & 0&06 & 27 & 1359 \\ \hline
 4& 6 &  5&0 & 0&43 & 22 &  638 \\
 6& 8 &  7&0 & 0&30 & 23 &  530 \\
 8&11 &  9&5 & 0&21 & 26 &  737 \\
11&15 & 13&0 & 0&13 & 26 &  778 \\
15&19 & 16&9 & 0&08 & 27 &  543 \\
19&24 & 21&3 & 0&06 & 27 &  468 \\
24&36 & 29&1 & 0&03 & 27 &  562 \\
\end{tabular}
\caption{Corrected average $M_X$, $\beta$ and $Q^2$ and the number of
observed events for the different intervals in $M_X$ considered in
this analysis.} 
\label{tab:intervals}
\end{center}
\end{table}

\subsection{Multiplicity moments}

The lowest-order moments of the multiplicity distribution, the
average multiplicity $\avg{n}$, the dispersion
$D=\avg{(n-\avg{n})^2}^{1/2}$
and the normalised second-order factorial moment $R_2$, have been
measured as a function of $M_X$.  The moment $R_2$ is defined as
$R_2 = \tilde{R}_2 / \avg{n}^2$ with
\begin{equation}
\tilde{R}_2 = \avg{n(n-1)} = \sum_n P_n \, n(n-1).
\end{equation}
\noindent The latter quantity is equal to the integral of the
inclusive two-particle density over a given domain in phase space
and is a measure of the strength of  correlations among
the produced hadrons~\cite{review:93,bib:h1_multiplicity}. Note that in
the
case of uncorrelated particle production, the probability to
produce $n$  particles, $P_n$,   follows a Poisson distribution with
the result that $R_2\equiv1$.

\begin{figure}
\epsfig{file=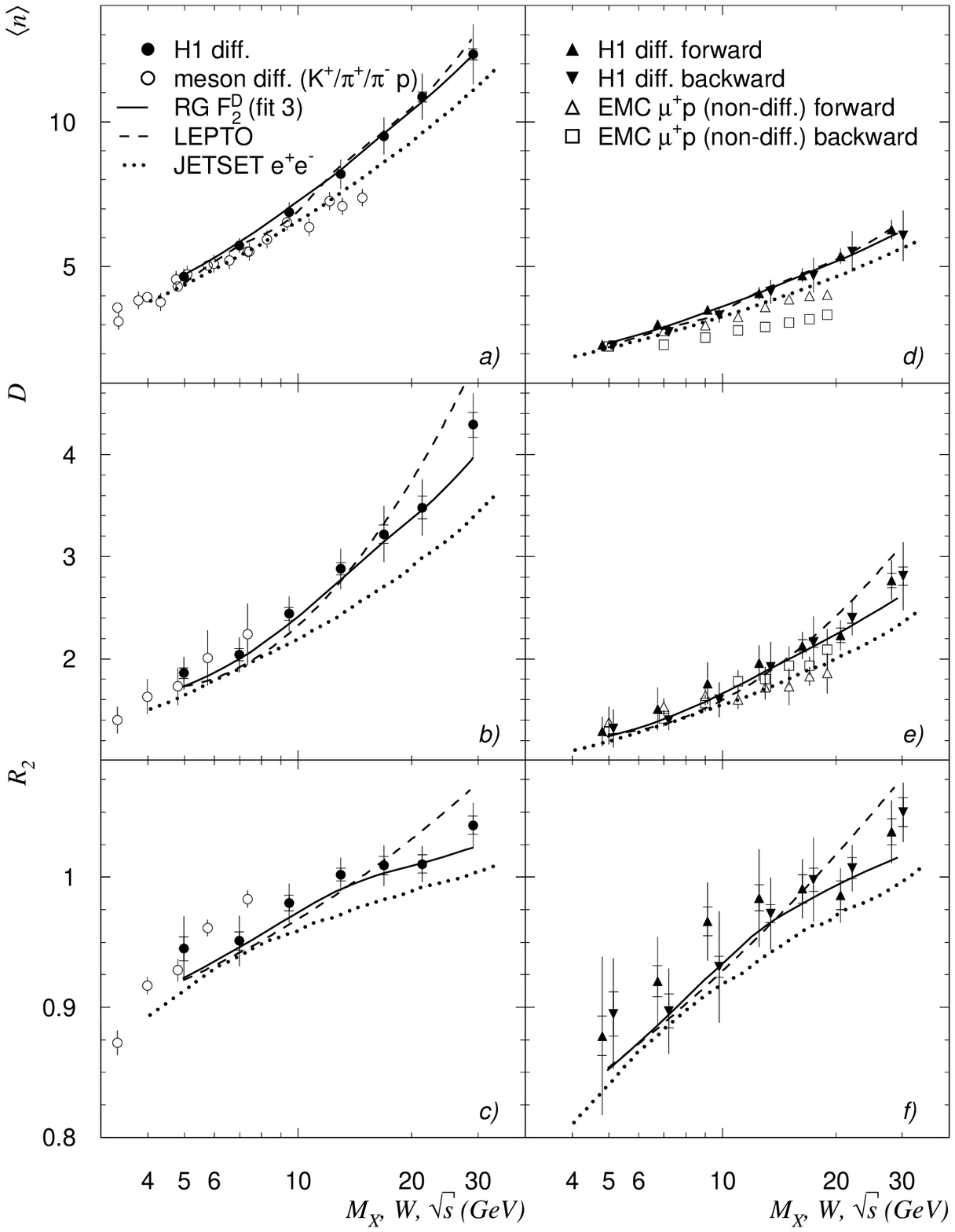}

\caption{Multiplicity moments $\avg{n}$, $D$ and $R_2$ in full
phase space {\bf(a-c)} and in single hemispheres {\bf(d-f)} 
for charged hadrons as a function
of $M_X$ (H1), $M_X$ (meson diffraction), $W$ (EMC), $\sqrt{s}$
($e^+e^-$), respectively.  For clarity, H1 data points in
single hemispheres are slightly shifted in the horizontal
direction with respect to their true positions.  Also shown are
predictions  of several Monte Carlo models (see text). The Monte
Carlo curves in forward and backward hemispheres are symmetric and their
average is plotted.}

\label{fig1} 
\end{figure}

Fig.~\ref{fig1}a shows the dependence of the mean charged
particle multiplicity
on  $M_X$ in full phase space. The H1 LRG data can be parameterized by a
form
$\avg{n} = a_1 + a_2 \ln M_X^2 + a_3 \ln^2 M_X^2$, 
with $a_1= 2.2 \pm0.4$, $a_2 =0.08 \pm 0.17$ and $a_3 = 0.21 \pm 0.02$ 
($\chi^2$ per degree of freedom $=0.4$; statistical errors only),
indicating that $\avg{n}$
increases faster
than the logarithm of the centre-of-mass energy of the system. In
non-diffractive DIS at HERA, a similar rate of increase 
is observed with respect to $W$~\cite{bib:h1_multiplicity}.
Also shown are measurements of $\avg{n}$ for the diffractively produced
system $X$ in the reactions  $\pi^\pm p\rightarrow X^\pm p$ and $K^+
p\rightarrow X^+ p$~\cite{bib:na22_diffmult,winkelmann}. Although the
two data sets agree well for $M_X\lsim10$~GeV, $\avg{n}$
in LRG events exceeds that in meson diffraction
at larger masses. There are no meson diffraction results with
$M_X\gsim15$~GeV. 

The meson-diffraction data are close to the $e^+e^-$ annihilation
results, represented here by the predictions from the JETSET parton
shower model (dotted line) which is known to reproduce well the $e^+e^-$
multiplicity data over a wide energy range (see
e.g.~\cite{bib:delphi_mdcorr,bib:opal_mdcorr}).

Results on the  dispersion and the correlation parameter $R_2$,
(Figs.~\ref{fig1}b,c) confirm that also the second-order moments in LRG
data are    similar  to meson diffraction 
and   $e^+e^-$  for $M_X \lsim 10$~GeV within the precision of the
measurements\footnote{The errors on the NA22 data points for $R_2$ are
derived from
published results for $\avg{n}$ and $\avg{n(n-1)}-\avg{n}^2$ whereby the 
(unknown) correlation between these quantities is neglected. The
quoted errors
are therefore to be considered as a lower limit on the true error.}.
Stronger multiplicity fluctuations and correlations 
than in $e^+e^-$  are observed at larger $M_X$.
The  rise of $R_2$ with $M_X$  shows
that KNO-scaling~\cite{bib:th_kno} does not hold in the $M_X$ range
studied here.

The similarities seen in Fig.~\ref{fig1}a-c 
have  led to the view~\cite{bib:misra,bib:na22_diffmult}
that in meson  diffraction  the (mainly longitudinal) momentum
exchange with the target leads to an excited meson state which can
be pictured as 
a colour-string of invariant mass $M_X$  stretched between
the  valence $q$ and $\bar q$ of the meson. This string subsequently
hadronises in a   way similar to a quark pair in $e^+e^-\rightarrow q\bar
q$ at the
corresponding centre-of-mass energy $\sqrt{s}=M_X$.
A comparative study of the thrust distribution, energy and quantum number
flow  in the rest-frame of the system $X$
and also in
 $e^+e^-$ further support  this
interpretation~\cite{bib:single:diffr}. 
For  $M_X\lsim10$~GeV
the same idea has been succesfully applied to proton
diffractive dissociation  assuming  that the
baryonic system $X$ now results
from the fragmentation of a (valence) quark-diquark string, 
thus explaining observed similarities with DIS lepton-nucleon
data at values of $W$ comparable to 
$M_X$~\cite{bib:na22_diffmult,pp:diffr}.
Due to the larger values of Bjorken-$x$ involved, the latter 
reaction  is  dominated by  quark-diquark fragmentation.

Combining these experimental results, it follows that low-mass
diffraction (above the resonance region) in hadron collisions and
photon dissociation in DIS at HERA may   be interpreted as
the hadronisation of a single string, or colour dipole, 
with colour triplet-antitriplet endpoints.
The larger  multiplicity moments seen in LRG events for  $M_X\gsim10$~GeV
relative to the other processes suggest, however, that 
the above interpretation is incomplete  and that high-mass diffraction
involves additional mechanisms.
Indications from experiment on the possible nature of these mechanisms
exist for hadron diffraction. 

In high-mass proton diffraction, measurements show that
the multiplicity structure of the system $X$ 
deviates from the expectations for quark-diquark fragmentation
and becomes, in fact,  similar to
that of soft non-diffractive interactions 
with $\sqrt{s}=M_X$~\cite{ua4:barnard}.
 
Within the  framework of the Dual Parton Model (DPM),  which is
phenomenologically very successful~\cite{bib:th_dpm}, soft
non-diffractive collisions are described by the fragmentation of
two or more strings corresponding to single or multiple pomeron
exchange in the elastic channel.
The similarity between non-diffractive and diffractive processes
is explained in DPM~\cite{DPM:DD} by assuming that  the colourless
exchange
in the latter becomes
resolved in a $q\overline{q}$ pair at large $M_X$
and  subsequently interacts with the
dissociating hadron.
The diffractive state  is then described by two colour
strings, one
stretched between a valence quark of the excited hadron
and a quark in the exchange, the other between
a diquark and the remaining quark of the  $q\overline{q}$ pair.

Multi-string systems are known to lead to a
faster than logarithmic increase of $\avg{n}$ with energy, to
wider multiplicity distributions and stronger long-range particle
correlations than  single-string  fragmentation~\cite{bib:th_dpm}. 
It is also important to note that only colour triplet-antitriplet strings
are considered in DPM. 


%

Additional mechanisms, besides $q\bar q$ fragmentation are included
in present models for diffractive DIS.
 Fig.~\ref{fig1}a-c show model calculations with RAPGAP~(fit~3) 
 (solid line) which describe the data well.
The difference between  RAPGAP and  JETSET $e^+e^-$ follows from
the presence, in RAPGAP, of additional diagrams involving
gluons from
the colourless exchange, leading to a large contribition
from boson-gluon fusion. The partonic state in lowest-order BGF
consists of a gluon (the ``pomeron remnant'') and a  $q\overline{q}$
pair in a colour-octet state. 
The fragmentation of this state allows for various string topologies,
including two-string configurations, thus leading one to expect
further  similarities with large $M_X$   hadron dissociation.

The admixture of the BGF sub-process with the $q\bar q$ and  QCD-Compton 
processes naturally explains 
larger mean multiplicity and stronger fluctuations. 
On the other hand, the results for RAPGAP~(fit~1), with a quark-dominated
pomeron
leading dominantly to $q\bar q$ parton states, are very similar to those
of JETSET~$e^+e^-$, as expected, and are not shown.

The data can also be qualitatively understood in the photon dissociation
picture of diffraction.
The lowest-order (``aligned jet''~\cite{bjorken:kogut})
excitation ($\gstar\rightarrow q\bar q$) is dominant for
$M_X^2<Q^2$ and leads to a  final state
similar to that in $e^+e^-$ annihilation.  In addition, higher-order
fluctuations (such
as
$\gstar\rightarrow q\bar q g$ where the gluon has low momentum),
which resemble the BGF sub-process, are believed to contribute at larger
$M_X$ and to
effectively interact as an octet-octet colour 
dipole~\cite{nik:zak:91,wusthoff:rapgap}. Due to the octet colour charge
at the dipole end-points, such a system hadronises 
with a larger mean multiplicity than a  $q\bar q$ 
state~\cite{opal:gluon:mul}. 

The LEPTO model with ``soft colour interactions'' (dashed curve) 
is seen  to also agree with the H1 data although it
predicts somewhat larger
multiplicity fluctuations above $M_X \sim 20$ GeV. This model   too
contains a sizeable BGF contribution.
However, diffraction is viewed here as a final-state interaction and does
not invoke   the notion of  colour-neutral exchange.
%


The moments of the multiplicity distribution for particles with
positive and negative rapidity (``forward'' and ``backward'',
respectively) are displayed in Figs.~\ref{fig1}d-f.  Rapidity
is calculated in the rest-frame of the system $X$ (the ``$\gstar\pom$''
centre-of-mass system) with
the positive
longitudinal momentum axis pointing in the $\gstar$
direction\footnote{The pomeron direction cannot be unambiguously
determined since the outgoing system $Y$ is not measured. Since
its transverse momentum
is small, it has been assumed that the $\pom$ direction is
collinear with the incident proton in the rest frame of the system
$X$.}, assigning the pion mass to  each charged track. 

The H1 data show no evidence for an
asymmetry between forward and backward hemispheres, in contrast to
what is observed for the  mean multiplicity  measured in
fixed-target  $\mu p$ DIS ($Q^2 > 4\,{\rm GeV}^2$)~\cite{bib:emc_mdcorr}, 
where the influence of proton
fragmentation on the backward-hemisphere  multiplicity distribution is
known to be substantial.


The $\mu p$ data in the current fragmentation region, where the
comparison with LRG data is most relevant, agree well with the
predictions for $e^+e^-$, as expected for production dominated by
quark jets. The LRG results are also here characterized
by larger $\avg{n}$ and stronger fluctuations above $M_X\gsim10$~GeV.

The RAPGAP and LEPTO models also predict
forward-backward
symmetry of the single-hemisphere moments and describe the data
adequately.  The earlier noted differences with $e^+e^-$
annihilation for the full phase space moments are also seen here.

The role of gluons in high-mass photon dissociation is prominent
in all the DIS  models considered, in contrast to  models for
hadronic soft diffraction where quark (diquark) fragmentation
is dominant. The indication from the models that the
pomeron-remnant has a large gluon content also opens interesting
opportunities for comparison with gluon-jet fragmentation in other
processes.

\subsection{Multiplicity distributions}

\begin{figure}
\epsfig{file=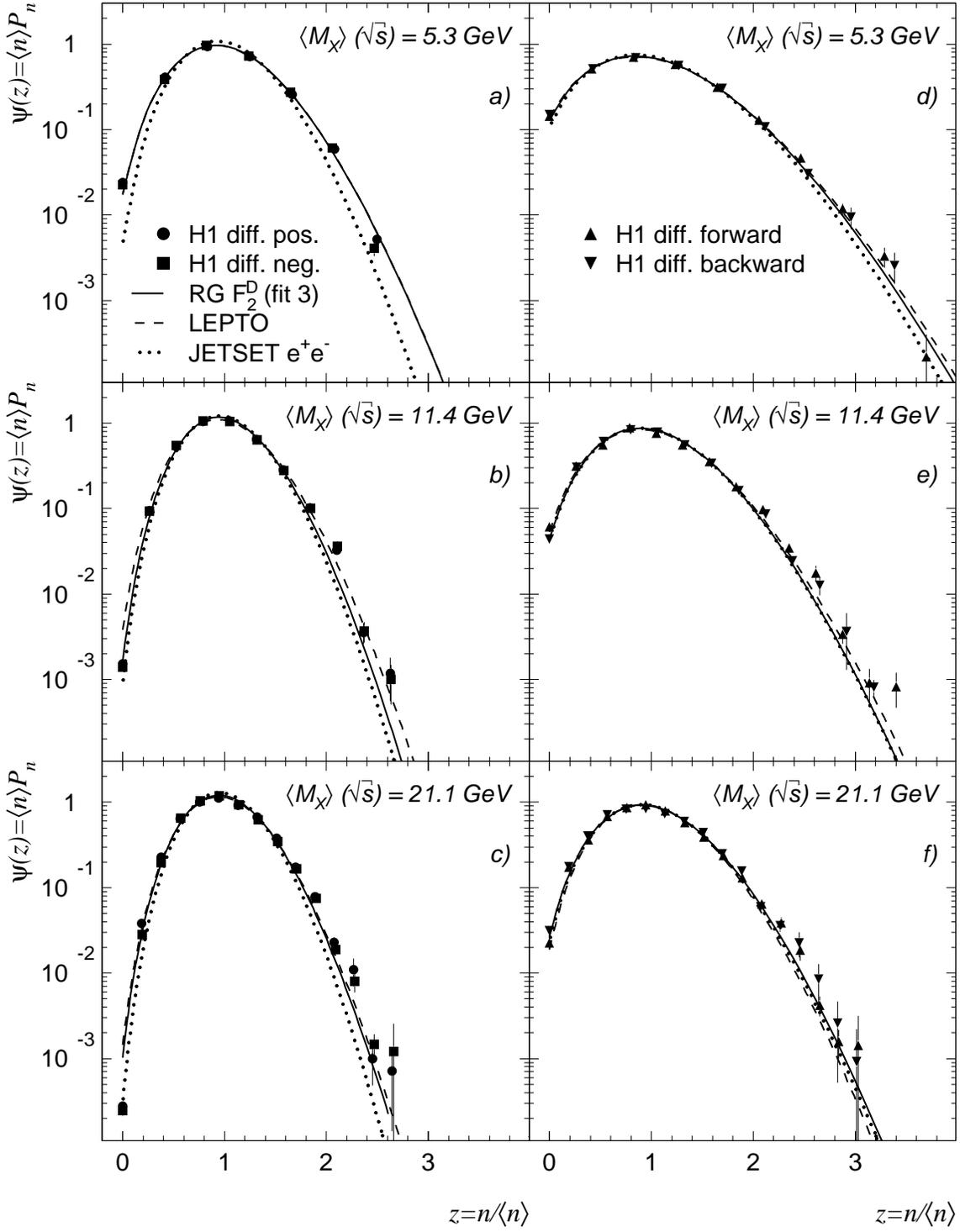}

\caption{The multiplicity distribution in KNO form for
three intervals in $M_X$ (DIS data and Monte Carlo) and at
$\sqrt{s}=\avg{M_X}$ (JETSET $e^+e^-$), in full phase
space for positive and negative particles separately {\bf(a-c)}, and
for all charges in single hemispheres {\bf(d-f)}.  The error bars show
statistical errors only.  
Also shown are
predictions  of several Monte Carlo models (see text).
The Monte Carlo curves are charge and
forward-backward symmetric  and their average is plotted.
}

\label{fig2}
\end{figure}

The multiplicity distributions in full phase space have been
measured, separately for negatively and positively charged tracks,
in three intervals of $M_X$.  The
results are displayed in Fig.~\ref{fig2}a-c in the form of a
KNO-distribution~\cite{bib:th_kno}:  $\psi(z) = \langle n\rangle
P_n$ plotted as a function of the normalised multiplicity $z = n /
\langle n\rangle$. No significant difference is observed between
the distributions for positively and negatively charged hadrons.
The data are well reproduced by the RAPGAP
model (solid curves) although there are indications that it
underestimates the high-multiplicity tail of the distribution at
large $M_X$. The
comparison with JETSET~$e^+e^-$ predictions (at $\sqrt{s}=\avg{M_X}$)
shows that the multiplicity distribution is broader in the LRG data,
indicative of stronger correlations among the hadrons. The predictions for
LEPTO are  similar to those of RAPGAP.

Figs.~\ref{fig2}d-f further illustrate the forward-backward symmetry
of the system $X$, now for  the all-charged multiplicity distribution. 
The RAPGAP and LEPTO predictions are also forward-backward symmetric 
but tend to fall below the data at large $z$. 
The  single-hemisphere distributions are closer
to the $e^+e^-$ expectations than in full phase
space (cfr.~Figs.~\ref{fig2}a-c). This difference can be understood as
the effect of correlations between hadrons emitted in opposite
hemispheres which, as will be shown below, are larger in the DIS LRG
data.

\subsection{Rapidity spectra}

\begin{figure}
\epsfig{file=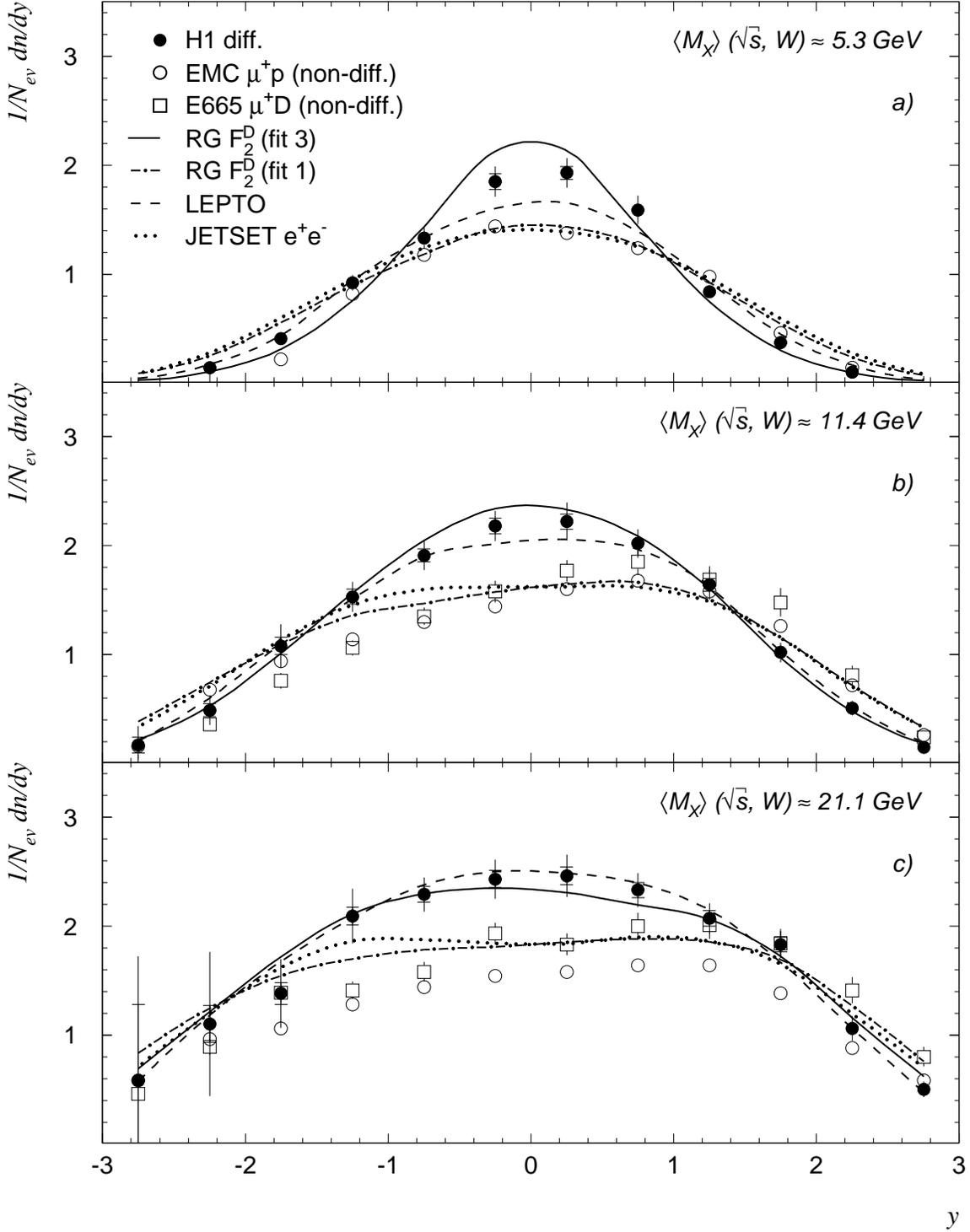}

\caption{Charged  particle rapidity spectra  for
three intervals in $M_X$ (H1), at $W=\avg{M_X}$ ($\mu N$) and at
$\sqrt{s}=\avg{M_X}$ (JETSET $e^+e^-$).
The $\avg{W}$ values for EMC and E665 differ slightly from the
ones indicated for H1 and are equal to 5.2, 11 and 19 GeV and 11.4
and 23.6 GeV, respectively. Also shown are
predictions  of several  Monte Carlo models (see text).}

\label{fig3}
\end{figure}

The charged particle rapidity density  in three intervals of $M_X$
is shown in Fig.~\ref{fig3}. The spectrum rises slowly with
$M_X$ in the central region and a rapidity plateau develops
with increasing phase space. These
features confirm earlier observations in hadron
diffraction~\cite{ua5:diffr,ua4:barnard,bib:single:diffr}
that the diffractive system
hadronises in a jet-like manner both in the forward and backward
regions~\cite{bib:h1:thrust,bib:zeus_eventshape,bib:h1:finalstates}. 
There
is no evidence for a significant forward-backward asymmetry of the
$y$-spectra\footnote{The rapidity spectra have also been
recomputed with rapidity defined along the thrust axis in the CMS
of the system $X$ (not shown). Except for the lowest $M_X$
interval, no significant difference is seen with the results in
Fig.~\ref{fig3}.  This is consistent with the observation
in~\cite{bib:h1:thrust} that the thrust axis in LRG events is
strongly aligned with the $\gstar$ direction in the $\gstar\pom$
system.} contrary to what is observed in $\mu N$
interactions~\cite{emc_comp,bib:e665}.

The particle density in the central region is much larger in  LRG
events than in $\mu N$ interactions
for  ${W}$ values close to $\avg{M_X}$. It is
also larger than in $e^+e^-$ annihilation (at $\sqrt{s}=\avg{M_X}$) 
according to the JETSET
expectation.  The RAPGAP~(fit~1) model curve  is close to JETSET and the
fixed-target data and  predicts a particle density which is too low.
Both RAPGAP with a hard gluon distribution and
LEPTO describe the rapidity spectra, although there are small 
deviations in the lowest $M_X$ bin. 

\begin{figure}
\epsfig{file=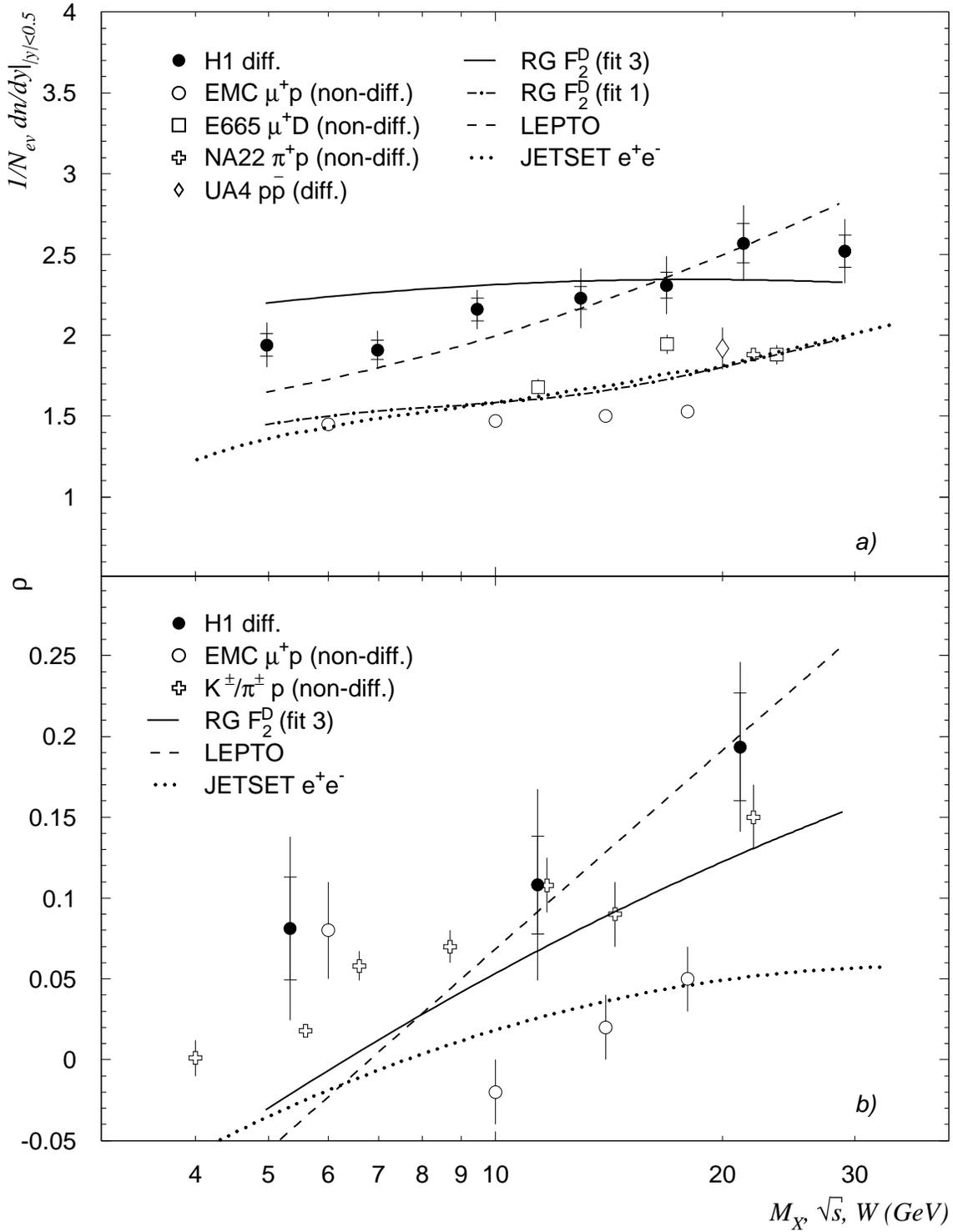}

\caption{{\bf(a)} Central region charged particle density ($-0.5
<y<0.5$); {\bf(b)} The
parameter $\rho$ (H1) and $b$ (others), which reflects the correlation
between the
number of particles in the forward and backward hemispheres, as a
function of $M_X$ (H1), $W=\avg{M_X}$ ($\mu p$), $M_X$ (hadron
diffraction), $\sqrt{s}$ (non-diffractive hadron-hadron),
$\sqrt{s}$ (JETSET $e^+e^-$). Also shown are
predictions  of several DIS Monte Carlo models (see text).}

\label{fig4}
\end{figure}

Fig.~\ref{fig4}a further compares the $M_X$ dependence of the central
particle density 
(defined as the mean multiplicity in the region $-0.5<y<0.5$) 
in LRG events 
to $e^+e^-$ expectations, to that in $\mu N$ collisions
\cite{emc_comp,bib:e665}, non-diffractive
meson-proton~\cite{bib:na22_rap}  collisions 
and proton diffraction~\cite{bib:ua4} .
The particle density near $y=0$ is seen to be 
larger in LRG events than in all the other processes.

The excess particle production relative to that in $e^+e^-$ and $\mu N$
indicates that additional mechanisms besides
hard and soft gluon bremsstrahlung from quarks are needed (cfr.
Sect.~3.1). 

The comparison with non-diffractive meson-proton  and
high-mass proton diffraction further shows that the
central particle density in
processes which are believed~\cite{bib:th_dpm} to involve
two or more strings with  colour triplet-antitriplet end-points
($q\bar q$ and quark-diquark strings) is  also significantly lower than in
the LRG data. 
This, together with previous observations, 
argues in favour of models which attribute  a higher gluonic
content to the partonic system created in 
virtual photon dissociation than in the other processes.

An estimate of the importance of  an
additional gluonic component may be obtained by assuming that the particle
density in the central region is a linear superposition of two
contributions, one arising from $q\bar q$
fragmentation (including additional QCD radiation), the second from a
colour octet-octet string
configuration.  This hypothesis is in line with expectations from
the photon dissociation picture of diffractive DIS
(see e.g.~\cite{wusthoff:rapgap}).

Using the EMC data for the former, and JETSET simulations
of a colour-singlet 
gluon-gluon string for the latter, it is found that, at
$\avg{M_X}=11.4$~GeV, 
approximately equal contributions of the two components
are needed to explain the particle density at 
mid-rapidity\footnote{This  is consistent
with the contribution of about $50\%$ from boson-gluon fusion events to
the
total diffractive cross section estimated with the RAPGAP model;
the latter value depends, however, on the 
$\hat{p_t}^2$ cut-off value, here chosen to be
2~GeV${}^2$~\cite{bib:h1:thrust}.}.

The RAPGAP (fit 3) predictions for the central particle density are shown
in
Fig.~\ref{fig4}a. They are compatible with the LRG data only above
$M_X\geq10$~GeV and are nearly $M_X$ independent.
 The LEPTO model, on the other hand, 
predicts a rather stronger  dependence on $M_X$, closer
to the tendency observed in the H1 data. 
The enhanced particle density, both in RAPGAP (fit 3) and in LEPTO,
are related to the large contribution from boson-gluon fusion.
The RAPGAP (fit 1) model version follows closely the JETSET $e^+e^-$
prediction.

In order to investigate the sensitivity of the results to
possible contributions from non-diffractive processes,
the analysis has been repeated changing the cut $\xpom<0.05$ to 
$\xpom<0.025$. 
Within errors, no significant effect on the results was observed.


\subsection{Forward-Backward Correlations}

In this section,  differences between LRG events and final
states in other processes are further examined through a
measurement of the correlation between hadrons emitted in opposite
event hemispheres.  These so-called ``forward-backward'' correlations
are known to be sensitive to finer details of the fragmentation
process and, in particular, to the presence 
in an inclusive event sample of several
distinct sub-classes of events\cite{bib:th_fbcorr}.


In previous experiments, the forward-backward correlation was analysed by
studying
the regression between the forward multiplicity, $n_F$, and the
backward multiplicity, $n_B$.
The correlation is usually well parameterised by a simple linear
dependence

\begin{equation} 
\aver{n_F}=a + b\cdot n_B.
\label{f:fb} 
\end{equation}

Forward-backward correlations have
not previously been measured in diffractively produced final
states.  
For reasons of statistics, matrix  techniques as used in
other analyses~\cite{bib:tasso_mdcorr,bib:delphi_mdcorr,bib:opal_mdcorr}
to unfold the two-dimensional forward and backward multiplicity
distributions
have not been employed.
 Instead, the 
forward-backward correlation
parameter is estimated from the separately unfolded and corrected
multiplicity distributions in full phase space, in the forward and
in the backward hemispheres.  Exploiting the relation between the
dispersion for the full phase space ($D$) and that for the forward and
backward hemispheres ($D_F$ and $D_B$), one can define the
correlation parameter $\rho$ as

\begin{equation} 
\rho = {D^2 - D_F^2 - D_B^2 \over 2 D_F D_B} . 
\label{eq:rho} 
\end{equation} 

\noindent The parameter $\rho$ is identical to the slope $b$ in
eqn.~(\ref{f:fb}) in the case of
forward-backward symmetric systems~\cite{bib:na22_fbcorr}.

Fig.~\ref{fig4}b shows the parameter $\rho$ in three intervals
of $M_X$ for the LRG data.  Also shown are data on the parameter $b$ for
$\mu p$ collisions~\cite{bib:emc_mdcorr},
for non-diffractive $\pi^\pm/K^\pm p$ collisions  compiled
in~\cite{bib:na22_fbcorr} 
and JETSET predictions  for $e^+e^-$.

In spite of the large errors,
there is clear evidence for stronger correlation in  LRG events than
observed in
 $e^+e^-$ annihilation~\cite{bib:tasso_mdcorr} and in the $\mu
p$ data for
energies above $\gsim10$~GeV. At lower energy, phase space effects
are important and mask possible differences in dynamics.
The  correlation strength  in diffractive DIS is comparable to that
in  meson-proton  interactions.

At LEP, where a value of  $b\sim0.1$ is measured, OPAL  
finds that  
the small correlation observed in an
inclusive sample of $e^+e^-$ events is primarily due 
to the superposition 
of events with distinct numbers of jets
and, therefore, different
average charged multiplicity. Sub-classes
of $n$-jet events
($n\geq 2$) show no or even negative 
correlations~\cite{bib:opal_mdcorr,bib:delphi_mdcorr}.
In $\nu p$ and
$\bar\nu p$ reactions~\cite{graessler_mdcorr} 
no clear evidence for  correlations is observed.
These data therefore show
that forward-backward correlations are small
at energies where the production mechanism is
believed to be dominated by single-string $q\bar q$  or quark-diquark
fragmentation.

In contrast, abundant data from hadron
interactions, compiled in~\cite{bib:na22_fbcorr}, which cover the
range  $10\leq\sqrt{s}\leq900$~GeV,  show that
the correlation  increases logarithmically with energy, with
$b$ as large as $0.65\pm0.01$ at $\sqrt{s}=900$ GeV.  
The strength and energy dependence of the effect 
is attributed to
strong event-to-event fluctuations of the
particle density 
as occur e.g.~in the multi-string Dual-Parton model
due to fluctuations in the number of strings and
strings overlapping in phase space\cite{bib:th_fbcorr}.

The observation of forward-backward correlations
in LRG events  with a strength comparable to that in
soft hadron interactions adds further  support
to the view that  the inclusive sample of DIS LRG events
is  a mixture of states with distinct hadronisation properties.
To disentangle their precise nature and relative contribution,
more differential studies will be needed, however.

In present models for diffractive DIS,
distinct production processes are readily identified and related, either
to the differences in parton composition and absorption probability
of virtual photon Fock-states, or to hard quark- and gluon-initiated
interactions off a colourless exchange.
That a mixture of such contributions leads to significant
forward-backward correlations is demonstrated by the predictions
for the parameter $\rho$ from
RAPGAP-$F_2^D$~(fit 3)   (solid line) and LEPTO (dashed) which  are close
to the H1 data for $M_X\gsim10$~GeV.  
The large difference between these
DIS models and JETSET for $e^+e^-$ illustrates the sensitivity of this
correlation measure
to differences in the dynamics of these two processes. 


\section{Summary and conclusions}

The charged-particle multiplicity structure of large-rapidity-gap events
of the type $\gstar p\rightarrow X Y$ in deep-inelastic scattering at HERA
has been
measured.
The major fraction of these events is generally interpreted as
due to diffractive dissociation of the virtual photon on the proton,
$\gstar p\rightarrow X p$.

Multiplicity distributions, lower-order moments, rapidity spectra
and correlations between hadrons emitted in opposite  hemispheres
in the rest-frame of the system $X$ have been presented as a function
of the invariant mass $M_X$.

The data have been compared with 
$e^+e^-$ annihilation (at $\sqrt{s}= M_X$), lepton-nucleon  data
in a $W$ range comparable to the $M_X$-range in the H1 data,
with hadro-produced diffractive final states, and also with data
from 
non-diffractive hadron-hadron collisions at $\sqrt{s}\sim M_X$.
The main observations are the following.
\begin{itemize}
\item The  mean total charged particle multiplicity $\avg{n}$ is a
function of $M_X$
and increases proportionally to $\ln^2{M_X}$.
The inclusive rapidity spectrum is forward-backward symmetric in the
rest-frame of $X$. 
A plateau develops with increasing
$M_X$. Both  $\avg{n}$ (for  $M_X\gsim10$~GeV)  and the  particle density
near $y=0$ 
are larger than in DIS at comparable values of $W$, 
than in $e^+e^-$ annihilation at $\sqrt{s}=M_X$ and than in hadro-produced
diffractive final states. Furthermore, the particle density in this 
central region is also higher
than in non-diffractive collisions at  $\sqrt{s}=M_X$.
\item For  $M_X\gsim10$~GeV, multiplicity  fluctuations are  larger than
in $e^+e^-$ annihilation and than in the current fragmentation region of
lepton-nucleon  interactions  at comparable values of $\sqrt{s}$ and $W$,
respectively.
The forward-backward multiplicity correlations are also larger and
of comparable strength to those measured in hadron interactions at
$\sqrt{s}=M_X$.

\end{itemize}

The distinctive characteristics of large-rapidity-gap events mentioned
can be globally understood if it is assumed that the
photon dissociation  mechanism involves a mixture of
different partonic states wherein gluons play an increasingly
important role as $M_X$ increases.
A large contribution from gluon-rich  states is
also required to explain the steep rise with increasing 
$1/x_{Bj}$, at
fixed $Q^2$, 
of the diffractive as well as the total virtual-photon proton cross
section~\cite{class,nik:zak:91}.

Good agreement with the data is  achieved with  a model which
assumes that the
diffractive process is initiated by the interaction of a
point-like virtual photon with a gluon-dominated colour-singlet
object emitted from the proton, as is suggested by a perturbative
QCD-Regge analysis  based on DGLAP evolution of the diffractive
structure function.
However, some deviations 
are  seen in the large-$n$ tail of the multiplicity distribution at high
$M_X$.

A model with soft colour interactions which rearrange the colour topology
after a normal deep-inelastic  scattering also describes the data although
the multiplicity fluctuations are somewhat overestimated
for $M_X$ larger than about 20 GeV.

The present  analysis adds new  support for the conclusion, derived from 
studies of event shapes~\cite{bib:zeus_eventshape,bib:h1:thrust}
and from a study of energy flow and  single particle
momentum spectra~\cite{bib:h1:finalstates}
in large-rapidity-gap events in H1, that gluons play a
prominent role in deep-inelastic diffraction.


\section*{Acknowledgements}

We are grateful to the HERA machine group whose outstanding
efforts made this experiment possible.  We thank the engineers and
technicians for their work in constructing and now maintaining the
H1 detector, our funding agencies for financial support, the DESY
technical staff for continual assistance, and the DESY directorate
for the hospitality they extend to the non-DESY members of the
collaboration.


\bibliography{h1,mc,theory,experiment,other}


\end{document}